\documentclass[
prl, 
reprint, 
superscriptaddress,preprintnumbers, amsmath, amssymb, amsfonts, nofootinbib]
{revtex4-1}
\usepackage{graphicx}

\usepackage{color}
\usepackage{axodraw}
\usepackage{dcolumn}
\usepackage{bm}

\newcolumntype{d}[1]{D{.}{\cdot}{#1}}
\newcolumntype{.}{D{.}{.}{-1}}
\newcolumntype{,}{D{,}{,}{2}}

\begin{document}

\title{Hadronic light-by-light scattering contribution to the muon anomalous magnetic moment from lattice QCD}

\newcommand{\RBRC}{
  RIKEN BNL Research Center,
  Brookhaven National Laboratory,
  Upton, New York 11973,
  USA}

\newcommand{\UCONN}{
  Physics Department,
  University of Connecticut,
  Storrs, Connecticut 06269-3046,
  USA}

\newcommand{\NAGOYA}{
  Department of Physics,
  Nagoya University,
  Nagoya 464-8602,
  Japan}
  
  \newcommand{\NISHINA}{
  Nishina Center,
  RIKEN,
  Wako, Saitama 351-0198,
  Japan}

  \newcommand{\BNL}{
  Physics Department,
  Brookhaven National Laboratory,
  Upton, New York 11973,
  USA}

\author{Thomas Blum}
\affiliation{\UCONN}
\affiliation{\RBRC}

\author{Saumitra Chowdhury}
\affiliation{\UCONN}

\author{Masashi Hayakawa}
\affiliation{\NAGOYA}
\affiliation{\NISHINA}

\author{Taku Izubuchi}
\affiliation{\BNL}
\affiliation{\RBRC}

\date{\today}

\begin{abstract}
The form factor that yields the light-by-light scattering contribution to the muon anomalous magnetic moment is computed in lattice QCD+QED and QED. A non-perturbative treatment of QED is used and is checked against perturbation theory.
The hadronic contribution is calculated for unphysical quark and muon masses, and only the diagram with a single quark loop is computed. Statistically significant signals are obtained. Initial results appear promising, and the prospect for a complete calculation with physical masses and controlled errors is discussed. 
\end{abstract}

\maketitle

\section{Introduction}
 The muon anomaly $a_\mu$ provides one of the most stringent tests of
the standard model because it has been measured
to great accuracy (0.54\ ppm)~\cite{Bennett:2006fi}, 
and calculated to even better
precision~\cite{Aoyama:2012wk,Davier:2010nc,Hagiwara:2011af}.
 At present, the difference observed between the experimentally 
measured value and the standard model prediction 
ranges between $249~(87)\times 10^{-11}$ and $287~(80)\times 10^{-11}$,
or about 2.9 to 3.6 standard
deviations~\cite{Aoyama:2012wk,Davier:2010nc,Hagiwara:2011af}.
 In order to confirm such a difference, 
which then ought to be accounted for by
new physics,
new experiments are under preparation at Fermilab (E989) and J-PARC (E34), 
aiming for an accuracy of 0.14 ppm.
 This improvement in the experiments, however, will not be useful
unless the uncertainty in the theory is also reduced.

 Table \ref{tab:summary} summarizes the
contributions to $a_{\mu}$ 
from QED~\cite{Aoyama:2012wk}, 
electroweak (EW)~\cite{Czarnecki:2002nt}, 
and QCD sectors of the standard model. 
The uncertainty in the QCD contribution
saturates the theory error.
The precision of the leading-order (LO) hadronic vacuum polarization (HVP) contribution
requires sub-percent precision on QCD dynamics, reached using 
a dispersion relation and either the experimental production cross 
section for hadrons ($+ \gamma$) in $e^+ e^-$ collisions at low energy,
or the experimental hadronic decay rate of the $\tau$-lepton
with isospin breaking taken into account.
Meanwhile lattice QCD calculations of this quantity
are improving rapidly~\cite{Blum:2013qu}, 
and will provide an important crosscheck.

\begin{table}[bt]
\caption{The standard model contributions to the muon $g-2$, 
scaled by $10^{10}$;
the QED contribution up to $O(\alpha^5)$,
EW up to $O(\alpha^2)$,
and QCD up to $O(\alpha^3)$,
consisting of 
the leading-order (LO) HVP, the next-to-leading-order (NLO) HVP, and HLbL.
 For the LO HVP 
three results obtained 
without (the first two)
and with (the last) $\tau\to\rm hadrons$ are shown.
}
\begin{center}
\begin{tabular}{cl|.r}
\hline
\hline
QED
& &116\,584\,71.8\,951~(9)(19)(7)(77)
 & \cite{Aoyama:2012wk}\\
\hline
EW &  & 15.4~(2)
 & \cite{Czarnecki:2002nt}\\
\hline
QCD
& LO HVP 
 & 692.3~(4.2) & \cite{Davier:2010nc}\\
& 
   & 694.91~(3.72)~(2.10) 
   & \cite{Hagiwara:2011af}\\
& 
 & 701.5~(4.7)
 & \cite{Davier:2010nc}\\
& NLO HVP & -9.79~(9) 
 &\cite{Hagiwara:2006jt}\\
& HLbL & 10.5~(2.6) 
 & \cite{Prades:2009tw}\\
\hline
\hline
\end{tabular}
\end{center}
\label{tab:summary}
\end{table}

\begin{figure}[hbt]
\begin{center}
\includegraphics[width=0.18\textwidth]{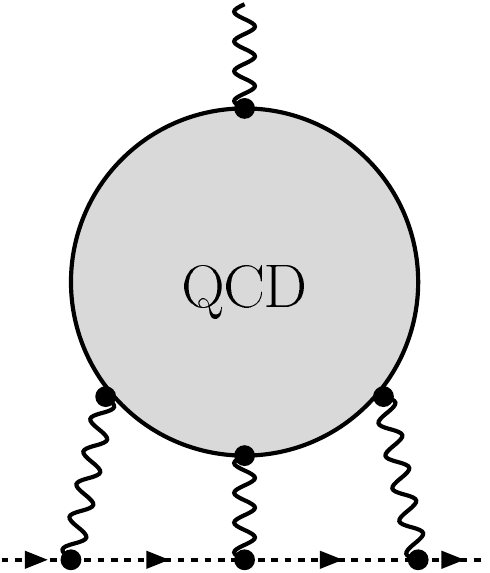} 
\caption{
 Hadronic light-by-light scattering contribution to the muon $g-2$, 
where the grey part consists of quarks and gluons.
 The wavy lines denote photons, 
and the dashed arrow line represents the muon.
}
\label{fig:HLbL}
\end{center}
\end{figure}

 Unlike the case for the HVP,
it is difficult, if not impossible, to determine the hadronic light-by-light scattering (HLbL) contribution (Fig.~\ref{fig:HLbL}),
$a_\mu(\rm HLbL)$,
 from experimental data and a dispersion relation~\cite{Colangelo:2014dfa}.
 So far, only model calculations have been done.
 The uncertainty quoted in Table \ref{tab:summary}
was estimated by the ``Glasgow consensus''~\cite{Prades:2009tw}.
 Note that the size of $a_\mu({\rm HLbL})$ is the same order as
the current discrepancy between theory and experiment.
 Thus, a first principles calculation,
which is systematically improvable, 
is strongly desired for $a_\mu({\rm HLbL})$.
In this paper, we present the first result for the magnetic form factor yielding $a_\mu({\rm HLbL})$ using lattice QCD.

\section{Non-perturbative QED method}\label{sec:method}
\begin{figure}[hbt]
\begin{center}
\includegraphics[width=.17\textwidth]{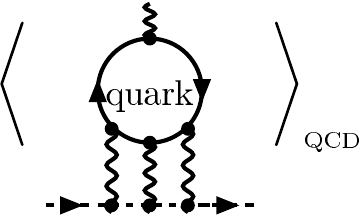} 
\includegraphics[width=.28\textwidth]{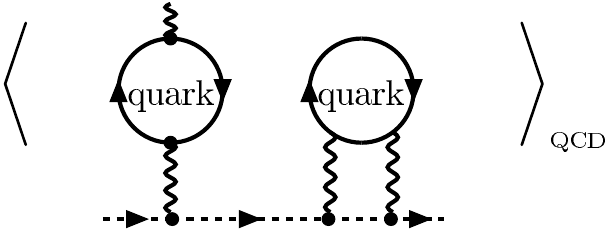} 
\caption{Two classes of diagrams contributing to 
$a_\mu({\rm HLbL})$.
 On the left, all QED vertices lie on a single quark loop,
 The right diagram is one of six diagrams where 
QED vertices are distributed over two (or three) quark loops. { We refer to these as (quark) connected and disconnected diagrams, respectively}.
}
\label{fig:classes hlbl diagrams}
\end{center}
\end{figure}

\vskip -0.15cm
 We start by observing the difficulty computing
$a_\mu({\rm HLbL})$ using lattice QCD, 
and then explain our strategy to overcome it. 
 Fig.~\ref{fig:classes hlbl diagrams} shows two (of seven) types of
diagrams, classified according to
how photons are attached to the quark loop(s).
 In the lattice calculation, the computation of
the vacuum expectation value 
of an operator involving quark fields requires
the inversion of the quark Dirac operator
$D_{m_q}\left[U^{\rm QCD}\right]$
for each gluon field (QCD configuration), $U^{\rm QCD}$.
 The cost of inversion of this operator
for every pair of source and sink points on the lattice is prohibitive
since it requires 
solving the linear equation $D_{m_q}\left[U^{\rm QCD}\right] x_r = b_r$
for $N_{\rm sites}$ number of sources, $b_r$, where $N_{\rm sites}$ is the total number of 
lattice points.
 In most problems, such as hadron spectroscopy,
all of these inversions are not necessary.
 For our problem, the correlation 
of four electromagnetic currents must be computed
for all possible values of two independent four-momenta.
 This implies $(3 \times 4 \times N_{\rm sites})^2$ separate 
inversions, per QCD configuration, quark species,
and four-momentum of the external photon 
to calculate the connected diagram 
in Fig.~\ref{fig:classes hlbl diagrams}, 
which is astronomical.
 Therefore, a practical method
with substantially less computational cost is indispensable.

 A non-perturbative QCD+QED method
which treats the photons and muon on the lattice
along with the quarks and gluons has been proposed as such a candidate by us.
 To obtain the result for the diagram in Fig.~\ref{fig:classes hlbl diagrams}
the following quantity is computed~\cite{Hayakawa:2005eq},
\begin{align}
 & \langle \psi (t',{\bf p'})\,j_\mu(t_{op},{\bf q})\,
\overline{\psi}
(0,{\bf p})\rangle_{\rm HLbL}=\nonumber\\
  & 
 - \sum_{q=u,d,s} \left(Q_q e\right)^2
   \sum_{{\bf y}} e^{-i {\bf q} \cdot {\bf y}}
   \frac{1}{L^3T}\sum_k \sum_x e^{-i k \cdot x} \sum_z e^{i k \cdot z}\nonumber\\
 & \quad
 \times
 \left\{
  \bigg<
   {\rm Tr}\{\gamma_\mu\,S_q(t_{op},\,{\bf y};\,z)\,\gamma_\nu\,
    S_q(z;\,t_{\rm op},\,{\bf y})\}
 \right.\nonumber\\
 &\qquad \quad 
  \left.
   \times
   \frac{\delta_{\nu\rho}}{\widehat{k}^{\,2}}\,
   G(t^\prime,\,{\bf p}^\prime;\,x) \gamma_\rho
   G(x;\,0,\,-{\bf p})
  \right\rangle_{\rm QCD + QED} \nonumber\\
 &\qquad 
  -
  \left<
  {\rm Tr} \{\gamma_\mu\,S_q(t_{op},\,{\bf y};\,z)\,\gamma_\nu\,
    S_q(z;\,t_{\rm op},\,{\bf y})\}
  \right>_{\rm QCD + QED} \nonumber\\
 &\qquad \quad 
 \left.
  \times
  \frac{\delta_{\nu\rho}}{\widehat{k}^{\,2}}\,
  \left<
   G(t^\prime,\,{\bf p}^\prime;\,x) \gamma_\rho
   G(x;\,0,\,-{\bf p})
  \right>_{\rm QED}
 \right\}\,.\label{eq:non pt method}
\end{align}
where $\psi$ annihilates the state with muon quantum numbers,
and $j_{\mu}$
is the electromagnetic current \footnote{The point-split, exactly conserved,
lattice current is used for the internal vertices
while the local current is inserted at the external vertex.}.
 $k$ is a 
{Euclidean four-momentum},
$\bf p$ is a three-momentum,
each quantized in units of $2\pi/L$.
 $\delta_{\mu\nu}/\hat k^{2}$
($\hat{k}_\mu \equiv 2 \sin({k_\mu/2})$)
is the momentum space lattice photon propagator in Feynman gauge. $L^3T$ is the space-time size of the lattice, $S_q$ and $G$ are quark and muon propagators, respectively,
and spin and color indices have been suppressed.
 One takes $t'\gg t_{op}\gg 0$ to project onto the muon ground state
\begin{eqnarray}
&&
\lim_{t'\gg t_{op}\gg 0}\langle
\psi (t',{\bf p'})\,j_\mu(t_{op},{\bf q})\,
\overline{\psi}
(0,{\bf p})\rangle_{\rm HLbL}= \nonumber\\
&&
\quad
\frac{\langle 0|\psi(0,{\bf p}^\prime)|{\bf p}^\prime,s^\prime\rangle}
     {2E^\prime V}
 \langle {\bf p^\prime},s^\prime|j_\mu|{\bf p},s\rangle
\frac{
\langle {\bf p},s| 
\overline{\psi}(0,{\bf p})
|0\rangle
}{2 E V} \nonumber\\
&&\quad\ 
\times
e^{-E^\prime(t'-t_{\rm op})}e^{-E t_{\rm op}},
\label{eq:lsz}
\end{eqnarray}
where the matrix element is parametrized, up to muon wave function renormalization factors, as
\begin{eqnarray}
 &&\langle {\bf p^\prime},s^\prime|j_\mu|{\bf p},s\rangle\equiv\\
 &&
\quad
-\bar u({\bf p'},s')\left(F_1(q^2)\gamma_\mu+i\frac{F_2(q^2)}{2m_{\mu}}\frac{[\gamma_{\mu},\gamma_{\nu}]}{2}q_\nu\right)u({\bf p},s).\nonumber
\label{eq:param}
\end{eqnarray}
$u({\bf p},s)$ is a Dirac spinor, and $q=p'-p$ is the space-like
four-momentum transferred by the photon. The minus sign in (\ref{eq:param}) results from the definition $F_1(0)=1$ and the fact that the muon has charge $-1$ in units of $e>0$.
To extract the form factors
$F_{1}$ and $F_{2}$,
Eq.~(\ref{eq:non pt method}) is traced over spins
after multiplication by one of the projectors,
$(1+\gamma_{t})/4$ or $i\, (1+\gamma_{t}) \gamma_{j}\gamma_{k}/4$,
where $j,k={x,y,z}$ and $k\neq j$.
 The contribution to the anomaly is then found
from $a_{\mu}\equiv(g_\mu-2)/2=F_{2}(0)$.

\begin{figure}[tp]
\begin{center}
\includegraphics[width=0.5\textwidth]{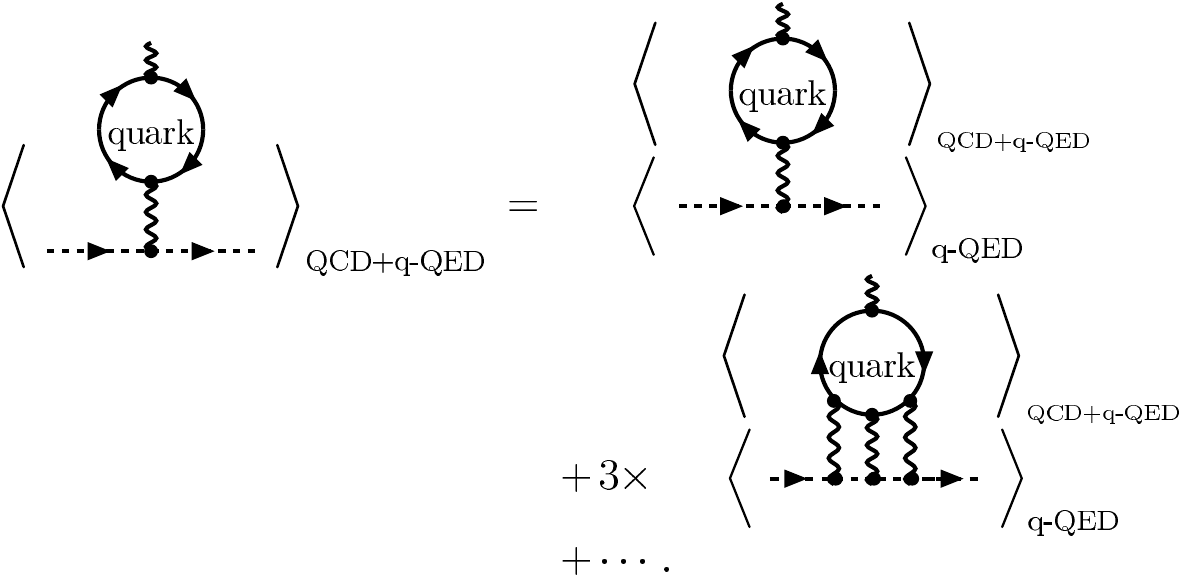} 
\caption{
 Perturbative expansion of
the first term in Eq.~(\ref{eq:non pt method}) with respect to QED.
 The symbols
$\left<,\,\right>_{{\rm QCD} + {\rm q\mathchar`-QED}}$
and $\left<,\,\right>_{{\rm q\mathchar`-QED}}$
represent
the average over QCD$+$QED configurations $(U^{\rm QCD},\,A^{\rm QED})$
and that over 
$A^{\rm QED}$, respectively.
 Terms represented by the ellipsis contain four
or more internal photons
{and so their orders are higher than $\alpha^3$}.
}
\label{fig:npt-method}
\end{center}
\end{figure}

 For now quenched QED (q-QED) is used for the QED average in (\ref{eq:non pt method}), 
implying no fermion-antifermion pair creation/annihilation via the photon.
 Note that only the sea quarks need to be charged under ${\rm U}(1)$;
the lepton vacuum polarization corresponds
to higher order contributions which we ignore.
 This approximation was chosen to make this first calculation
computationally easier, even though it is incomplete.
 We can remove it to get the complete physical result,
as discussed at the end of this letter.
 The first term, expanded in q-QED, 
can be reorganized
as in Fig.~\ref{fig:npt-method}, 
according to the number of photons exchanged between 
the quark loop and the open muon line. 
 If the second term in 
Eq.~(\ref{eq:non pt method}) is subtracted, 
the connected diagram in Fig.~\ref{fig:classes hlbl diagrams}, 
times $3$ (the multiplicity arises because two of the three internal photon lines are generated three ways), emerges as the leading-order contribution in $\alpha$.

 The main challenge in the non-perturbative method
is the subtraction of the leading, unwanted components ($\alpha$ for the electric form factor and $\alpha^{2}$ for the magnetic).
 Note that the two terms in 
Eq.~(\ref{eq:non pt method})
differ only by way of averaging. 
 For finite statistics, the delicate cancellation between them
is only realized because they are so highly correlated
with respect to the QCD and QED configurations used in the averaging. 
{Notice that all contributions from one-photon exchange between the lepton (quark) loop and muon line  are canceled by the subtraction. However, two photon exchange contributions, which vanish by Furry's theorem after averaging over gauge fields, cannot appear in the subtraction term and are a potential source of large statistical errors. Fortunately these too can be completely eliminated on each gauge configuration by switching the sign of the external momentum. This is because the projected and traced correlation function in (\ref{eq:non pt method}) obeys an exact symmetry under simultaneous $\bf p\to -p$ and $e\to-e$, where the latter is done on the muon line only. If $e$ does not flip sign, then the only change is to multiply all contributions with an even number of photons connecting the loop and line by $-1$.}

We first test the non-perturbative subtraction by asking if 
the nonperturbative QED method applied to leptons only
reproduces the known value of
the sixth-order leptonic light-by-light scattering contribution 
~\cite{Aldins},
which is given exactly by the counterpart of the connected diagram in 
Fig.~\ref{fig:classes hlbl diagrams}. 

\begin{figure}[htbp]
\begin{center}
\includegraphics[width=0.45\textwidth]{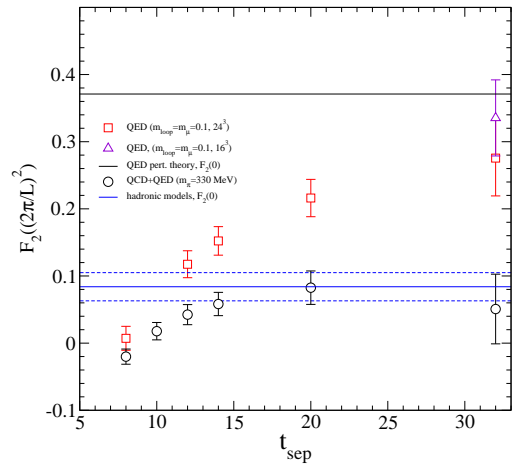} 
\caption{The muon's magnetic form factor in units of $(\alpha/\pi)^{3}$ from light-by-light scattering, evaluated at the lowest non-trivial lattice momentum, $2\pi/L$. Results for several symmetric source-sink separations are shown; the quark loop is the same for each and corresponds to $m_{\pi}=329$ MeV (circles). Also shown is the pure QED result (squares, triangle) where the mass of the lepton in the loop is equal to the muon mass, $m_{\mu}=0.1$. Horizontal lines correspond to continuum QED perturbation theory (upper) and hadronic models~\cite{Prades:2009tw} (lower). A large excited state contamination is evident for both cases.}
\label{fig:f2vstsep}
\end{center}
\end{figure}

{
 The test calculation was done
in quenched
\footnote{In the pure QED case,
quenching is not an approximation since the neglected vacuum polarization contributions give higher order corrections to the light-by-light scattering diagram.} non-compact QED, in the Feynman gauge,
using domain wall fermions (DWF). Non-compact  here only refers to the generation of the photon field configurations; the photons are coupled to the fermions via the usual exponentiated link variable.
 The lattice size is $24^{3}\times 64$ with $L_{s}=8$ sites
in the extra 5th dimension and domain wall height $M_5=0.99$.
 The muon mass and the lepton mass are the same, 0.1 in lattice units,
and to enhance the signal the electric charge is set to $e=1.0$,
which corresponds to $\alpha=1/(4\pi)$ instead of $1/137$.
 For simplicity, we always use kinematics where the incoming muon
is at rest. The form factor $F_{2}$ was computed
only at the lowest non-trivial momenta, $2 \pi/24$,
and was not extrapolated to zero.
 The renormalization factor of the local vector current
inserted at the external vertex is not included 
as its effect is $O(\alpha)$ and should be small compared
to other uncertainties. 

The results for several values of the time separation between the muon source and sink, $t_{\rm sep}$, are shown in Figure~\ref{fig:f2vstsep} (squares). The results were computed from an ensemble of 100 uncorrelated configurations and $6^{3}=216$ point source locations for the external photon vertex which was inserted on time slice $t_{\rm op}=5$. The form factor shows a large excited state contamination, presumably arising from the contribution of muon-photon states. The value for the largest separation ($t_{\rm sep}=32$) is still somewhat below the continuum result, $F_{2}(0) = 0.371 (\alpha/\pi)^{3}$~\cite{Laporta}. There may be residual excited state contamination, or finite volume and lattice spacing effects. A smaller $16^3$ lattice size result (triangle) is consistent, within errors, with the $24^3$ result, and the lattice result also must be extrapolated to $Q^2=0$ before the final comparison with continuum perturbation theory.
}

\section{QCD contribution}

 The inclusion of QCD into the light-by-light amplitude is straightforward:
simply construct combined links from the product of ${\rm U}(1)$ (QED) gauge links and ${\rm SU}(3)$ (QCD) links~\cite{Duncan:1996xy},
and follow exactly the same steps, using the same code,
as described in the previous sub-section.
 We use one quenched QED configuration per QCD configuration,
though different numbers of each could be beneficial and should be explored.

 Our main result is again computed on a lattice of size $24^{3}\times 64$ ($L_{s}=16$, $M_5=1.8$)
with spacing $a=0.114$ fm ($a^{-1}=1.73$ GeV)
and light quark mass 0.005 ($m_{\pi}=329$ MeV)
(an RBC/UKQCD collaboration 2+1 flavor,
DWF+Iwasaki ensemble~\cite{Allton:2008pn,Aoki:2010dy}).
 The bare muon mass is again set  to $m_{\mu}=0.1$ (the renormalized mass extracted from the two-point function is 190 MeV),
and $e=1$ as before. The domain wall height $M_5$ for the quark loop propagators is set to 1.8, the value used to generate the gluon gauge field ensemble; $M_5$ for the muon line is the same as in the pure QED case.

The all mode averaging (AMA) technique~\cite{Blum:2012uh}
is used to achieve large statistics at an affordable cost.
In the AMA procedure the expectation value of an operator is given by
$\langle {\cal O} \rangle = \langle {\cal O}_{\rm rest}\rangle+\frac{1}{N_G}\sum_{g}\langle {\cal O}_{\rm approx,g}\rangle$~\cite{Blum:2012uh},
where $N_G$ is the number of measurements of the approximate observable,
and ``rest" refers to the contribution of the exact observable
minus the approximation, evaluated for the same conditions.
The exact part of the AMA calculation was done
using eight point sources on each of 20 configurations, and
the approximation was computed using 400 low-modes of
the even-odd preconditioned Dirac operator and $N_G=216$ point sources computed with stopping residual $10^{-4}$
on $375$ configurations. On a different subset of 190 configurations
we tried 125 point sources and found the 216 sources per configuration to be more effective
at reducing the statistical error. In the present calculation, the statistical errors are completely dominated by the second term in the above equation, (approximately 4:1) and the ``rest", or correction is about $-10\pm5\%$.

The external electromagnetic vertex is inserted on time slice $t_{\rm op}=5$
with the muon created and destroyed at several time separations ranging between 8 and 20. 
We also include the vector current renormalization in pure QCD
from~\cite{Aoki:2010dy} for the local vector current at the external vertex. We have computed the connected diagram shown in Fig.~\ref{fig:classes hlbl diagrams} for a single quark with charge +1 in the present exploratory study, so the final result is multiplied by $(2/3)^4+(-1/3)^4+(-1/3)^4$ to account for (degenerate) $u$, $d$, and $s$ quark contributions.

In Fig.~\ref{fig:f2vstsep} we show $F_2((2\pi/L)^{2})$ for hadronic light-by-light scattering. Again there is a large excited state effect. For $t_{\rm sep}=20$ the ground state appears to dominate, and the value is roughly consistent with the model estimate~\cite{Prades:2009tw}. By $t_{\rm sep}=32$, the signal has disappeared, but there is no suggestion of large residual excited state contamination. The unphysical heavy masses used here for numerical expediency are expected to lead to a somewhat higher value: in hadronic models the increase due to muon mass overwhelms the decrease due to heavier pion mass~\cite{private}. 

\begin{figure}[htbp]
\begin{center}
\includegraphics[width=0.45\textwidth]{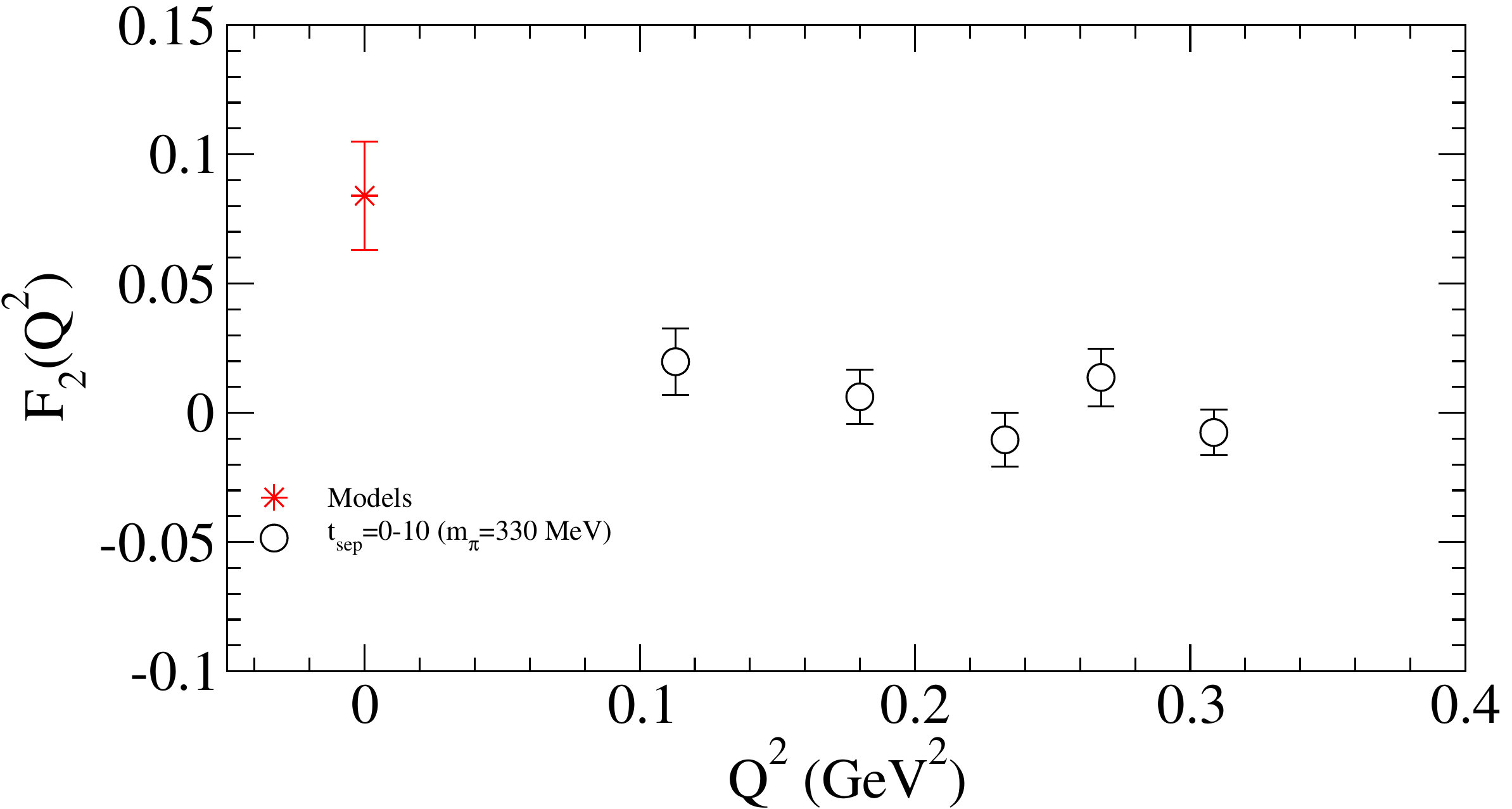} 
\caption{The muon's magnetic form factor in units of $(\alpha/\pi)^{3}$ from hadronic light-by-light scattering. $m_{\pi}=329$ MeV. The time separation between the muon source and sink in this case is $t_{\rm sep}=10$. The model result (burst) is for physical masses.}
\label{fig:f2vsqsq}
\end{center}
\end{figure}

$F_{2}(Q^{2})$ is shown in Fig.~\ref{fig:f2vsqsq} for several values of $Q^{2}$ for $t_{\rm sep}=10$. A mild dependence on $Q^{2}$ is seen. While we have not computed $Q^2$ values for $t_{\rm sep}=20$, a similar dependence is expected since the quark part computed in both is the same; only the muon line is different.

{As anticipated above, before averaging over equivalent external momenta,
the statistical errors are considerably larger as the two photon exchange contribution is one order lower in $\alpha$. While the combinations $\pm\vec p$ effectively eliminate the error from this contribution, the light-by-light contribution is identical, so the statistical error is only reduced by averaging over independent momenta or the $\gamma^\mu$ inserted at the external vertex. } 

Early preliminary work~\cite{Blum:2011pu} was done on another DWF+Iwasaki ensemble with size $16^{3}\times 32$ and light quark mass $m_{q}=0.01$ ($m_\pi=422$ MeV). Two muon masses,  $m_{\mu}=0.4$ and 0.1, were used. The external electromagnetic vertex is inserted on time slice $t_{\rm op}=6$
and the incoming and outing muons are created and destroyed at $t=0$ and 12,
respectively. Following the same procedure as above (except that we did not use AMA), for $m_{\mu}=0.4$ (6.5 times the physical muon mass), 
{$F_{2}(Q^2=0.38\,\rm GeV^{2})= (5.8\pm0.6) \times 10^{-5}=(0.79\pm0.08)\times(\alpha/\pi)^{3}$}. 
The magnitude is roughly 5 times larger than the model estimates for $a_{\mu}(\rm HLbL)$. The smaller muon mass $m_{\mu}=0.1$ yields 
 {$F_{2}(Q^2=0.19\,\rm GeV^{2})= (0.48\pm0.18) \times 10^{-5}=(0.065\pm0.024)\times(\alpha/\pi)^{3}$}.

 Finally, the subtraction is shown to be working properly
in the (QCD + QED) case by varying $e$
as follows.
 The same non-compact QED configurations are used in each case;
 $e$ is varied only when constructing the exponentiated gauge-link,
$U_{\mu}(x) = \exp\left(ie A_{\mu}(x)\right)$.
 Thus the ratio of form factors, and hence the $\alpha$-dependence,
can be determined very accurately.
 Since one photon is inserted explicitly, and 
the charges at the associated vertices are
not included in the lattice calculation, the subtracted amplitude
should behave like $e^{4}\propto\alpha^{2}$.
 Using $e=0.84$ and 1.19, the changes in the subtracted correlation function
relative to $e=1$ should be 0.5 and 2.0, respectively.
 This is what is observed numerically.

\section{Conclusion}
\label{sec:con}

 We have presented the first lattice QCD result 
for the form factor that yields the hadronic light-by-light contribution to the muon anomaly.
 The calculation uses a nonperturbative QED method
whose feasibility was first tested in the pure QED case. We have demonstrated that a statistically significant signal for the light-by-light diagram can be computed with modest statistics and that realistic results are obtained on modest size lattices. Large excited state contamination is visible in both QED and QED+QCD, likely attributable to the same muon+photon state. With large enough time separation between the muon source and sink, results for unphysical quark and lepton masses emerge that are consistent with expectations from model calculations and QED perturbation theory. A precise calculation with physical masses, larger volume, and a controlled extrapolation to $Q^2=0$, is now desirable and appears feasible.

 An additional {systematic uncertainty in the current calculation arises from
 the} absence of diagrams with two or more quark loops
coupled to photons like the one shown
on the right in Fig.~\ref{fig:classes hlbl diagrams}. This is a direct consequence of the numerical expediency of quenched QED in this first calculation.
The disconnected diagram in Fig.~\ref{fig:classes hlbl diagrams} (as well as the five others not shown) is next-to-leading order in the number of colors and
vanishes in the ${\rm SU(3)}$-flavor symmetry limit. We note that all such diagrams can be included in an analogous calculation to the one presented here, but using completely unquenched QED+QCD gauge field configurations~\cite{BlumLat2013}. These can be dynamical QED+QCD configurations~\cite{Horsley:2013qka,Borsanyi:2014jba}, or pure QCD ones, reweighted to non-zero electric charge~\cite{Ishikawa:2012ix}.

\section*{Acknowledgements}
{We thank
Norman Christ, Luchang Jin, and Christoph Lehner for useful discussions, and for help checking our code against independently written code, including PhySyHCAl\cite{Lehner:2012bt}.} We also thank the USQCD Collaboration and the RIKEN BNL Research Center for computing resources.
 MH is supported in part by Grants-in-Aid for Scientific Research
22224003, 25610053,
 TB is supported in part by the US Department of Energy
under Grant No. DE-FG02-92ER41989,
 and TI is supported in part by Grants-in-Aid for Scientific Research
22540301 and 23105715 and under U.S. DOE grant DE-AC02-98CH10886.

\end{document}